\documentclass[conference]{IEEEtran}
\IEEEoverridecommandlockouts

\usepackage{cite}
\usepackage{amsmath,amssymb,amsfonts}
\usepackage{algorithmic}
\usepackage{graphicx}
\usepackage{textcomp}
\usepackage{xcolor}
\usepackage{booktabs}
\usepackage{multirow}
\usepackage{siunitx}
\usepackage{url}

\renewcommand{\baselinestretch}{0.98}

\def\BibTeX{{\rm B\kern-.05em{\sc i\kern-.025em b}\kern-.08em
    T\kern-.1667em\lower.7ex\hbox{E}\kern-.125emX}}

\begin{document}

\title{Beyond Encoder Fusion: Multi-View Discrete Token Augmentation for LLM-Based ASR%
}

\author{
\IEEEauthorblockN{Paul Moïse Gangbadja}
\IEEEauthorblockA{
\textit{LIA, Avignon University, France}\\
\textit{EDL, France}\\
paul-moise.gangbadja@alumni.univ-avignon.fr}
\and
\IEEEauthorblockN{Mickael Rouvier}
\IEEEauthorblockA{
\textit{LIA, Avignon University, France}\\
mickael.rouvier@univ-avignon.fr}
\and
\IEEEauthorblockN{Fabrice Lefevre}
\IEEEauthorblockA{
\textit{LIA, Avignon University, France}\\
fabrice.lefevre@univ-avignon.fr}
}


\maketitle

\begin{abstract}
Discrete speech tokens provide a compact interface between speech encoders and large language models for automatic speech recognition, but single-tokenization systems remain sensitive to the chosen encoder. We propose multi-view discrete token augmentation, a simple strategy that augments each training utterance by generating alternative token sequences from fixed SSL encoders, such as HuBERT, WavLM, and MMS-300M. These tokenizations are treated as complementary training views for a shared LLM decoder, exposing it to more diverse discrete speech representations without requiring multi-encoder inference. At test time, the model can operate with a single encoder. On LibriSpeech, the approach consistently improves all encoders over independently trained baselines, with WavLM reaching 3.30\% WER on test-clean and 8.13\% on test-other. Budget-matched controls show that the gains come from encoder diversity rather than data volume. ROVER over multi-view hypotheses further improves WER to 3.03\% and 7.38\%.
\end{abstract}
 
\begin{IEEEkeywords}
automatic speech recognition, large language models, self-supervised
learning, discrete speech tokens, data augmentation, ROVER.
\end{IEEEkeywords}

\section{Introduction}
\label{sec:intro}

Speech-based large language models (LLMs) commonly rely on either continuous
or discrete speech representations. Continuous approaches project speech
encoder embeddings into the LLM input space, whereas discrete approaches
quantize speech into symbolic units that can be handled like text
tokens~\cite{wang2025discrete}. Discrete speech tokens are attractive because they integrate naturally with the LLM
vocabulary, shorten the input sequence, and reduce training cost. Despite these advantages, improving the accuracy of discrete-token systems remains an open challenge~\cite{wang2025discrete}. This paper focuses on this challenge within the discrete-token regime and asks how to better exploit SSL encoders for LLM-based ASR.


A promising way to improve these representations while preserving the efficiency of discrete tokens is to exploit the complementary information captured by different SSL encoders. Since these models are trained with different objectives and data, their representations can emphasize different speech properties. For example, HuBERT is known to capture strong phonetic structure~\cite{hsu2021hubert}, whereas WavLM incorporates noise robustness and speaker-related information through its pre-training design~\cite{chen2022wavlm}. Existing approaches usually merge encoder information before decoding, either by concatenating
embeddings before quantization or by combining token streams after
quantization. In this paper, we propose \emph{multi-view discrete token augmentation}: instead
of merging encoder outputs before decoding, we treat SSL encoders as
alternative discrete tokenizers of the same utterance. Each encoder produces an
encoder-specific tokenization, or \emph{view}, and a shared LLM decoder is
trained to map all views to the same transcript.

This approach is related to subword regularization~\cite{kudo2018subword} and
BPE-dropout~\cite{provilkov2020bpedropout}, which train text models with
multiple segmentations of the same sentence. Here, however, the alternative
tokenizations come from different pre-trained speech encoders, so they differ
by model bias rather than by random tokenization noise. A practical advantage
is that the extra encoders are needed only during training. At inference, the
decoder can be queried with a single encoder, keeping the cost of a standard
single-encoder system.

We evaluate this method on LibriSpeech~\cite{panayotov2015librispeech} and
Loquacious~\cite{parcollet2025loquacious}. Our experiments test whether
multi-view training improves each encoder over its single-encoder baseline.
They then examine whether the gains come from encoder diversity rather than
data volume alone. Finally, we assess whether the resulting hypotheses remain
complementary enough to be combined with ROVER~\cite{fiscus1997rover}.

Our contributions are:
\begin{itemize}
\item We propose \emph{multi-view discrete token augmentation}, where fixed SSL encoders provide alternative token views of the same utterances to train one shared LLM decoder.
\item We show that the gains come from view diversity, using repeated-encoder,
fixed-budget, and pairwise controls.
\item We compare against early and late fusion baselines and show that
multi-view token augmentation gives stronger LibriSpeech results at
single-encoder inference cost.
\item We show that encoder hypotheses remain complementary and can be further
improved with ROVER.
\end{itemize}

\section{Related Work}
\label{sec:related}


This section relates the proposed method to prior work on speech
representations for LLMs (Section~\ref{sec:rw_repr}), multi-encoder
speech systems (Section~\ref{sec:rw_multi}) and tokenization-based augmentation (Section~\ref{sec:rw_aug}).

\subsection{Discrete Speech Tokens and Tokenized Speech Modeling}
\label{sec:rw_repr}

Discrete-token approaches convert speech into symbolic sequences before
language-model decoding. A common pipeline extracts SSL features, quantizes
them with K-means or related methods, removes consecutive repetitions, and
optionally applies BPE to obtain more compact token sequences. These units can then be added to the LLM vocabulary and processed in a similar way to text tokens. This paradigm has been explored in systems such as SpeechGPT~\cite{zhang2023speechgpt}, AudioPaLM~\cite{rubenstein2023audiopalm} and discrete-token spoken language understanding models~\cite{shon2024discreteslu}.

The choice of SSL encoder is central in such systems, since the discrete
tokens inherit the biases and information content of the underlying
continuous speech features. Different encoders may encode phonetic, speaker,
language, or robustness-related information differently, depending on their
pre-training objectives and data. Most discrete-token systems rely on one
encoder and therefore on one tokenization of the speech signal. The method
proposed in this paper instead treats several SSL encoders as alternative
tokenizers of the same utterance, producing multiple discrete views that are
used to train a shared LLM decoder.


\subsection{Combining Multiple Speech Encoders}
\label{sec:rw_multi}

Several studies have investigated the use of multiple speech encoders to
exploit complementary information. FeaRLESS~\cite{chen2022fearless} showed
that naive concatenation or weighted sums of continuous SSL features may fail
because encoder representations can be highly correlated. Wang et
al.~\cite{wang2024fusion} fused two discrete SSL
streams with a cross-attention encoder in order to handle the misalignment caused by
de-duplication and BPE. Other recent
approaches exploit encoder diversity through routing or expert selection.
MoWE-Audio~\cite{zhang2025mowe} routes among speech encoders in the
continuous regime, while HDMoLE~\cite{mu2025hdmole} routes over LoRA experts
for accent adaptation. The proposed approach takes a different direction:
encoder-specific tokenizations are kept separate and used as alternative
training views of the same utterance. The encoders are therefore not fused at
the input level. Instead, they provide training-time diversity for a shared
decoder, while output-level combination can still be applied after decoding.


\subsection{Data Augmentation and Tokenization Diversity}
\label{sec:rw_aug}

Data augmentation is widely used in ASR, most often at the waveform or feature level through speed perturbation, additive noise, reverberation, or
SpecAugment~\cite{park2019specaugment}. These methods expose the model to
perturbed versions of the same training utterances, encouraging robustness to
acoustic variability. In text modeling, augmentation can
also act at the tokenization level: subword regularization
~\cite{kudo2018subword} and BPE-dropout~\cite{provilkov2020bpedropout}
train on several segmentations of the same sentence. Multi-view token augmentation transfers this principle to discrete speech
tokens. Instead of perturbing one tokenizer, we obtain
alternative discrete tokenizations from different SSL encoders. These views
differ by pre-training objective and model bias, and are used as
training-time augmentation for an LLM-ASR decoder.

\section{System Overview}
\label{sec:system}



\begin{figure}[htbp]
\centerline{\includegraphics[scale=0.10]{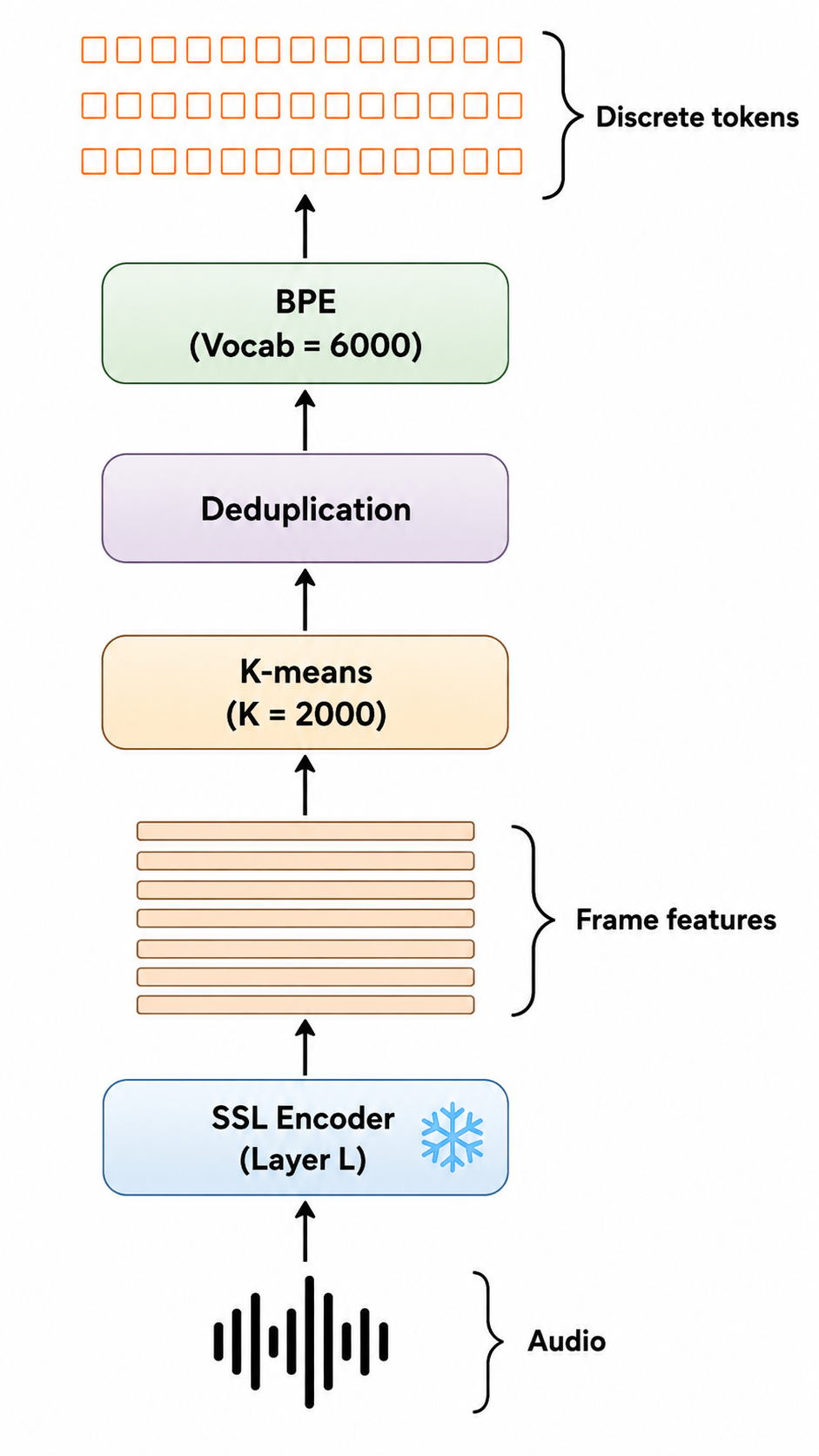}}
\caption{Discrete speech tokenization. A frozen SSL encoder
extracts frame-level features, which are quantized with K-means
($K{=}2000$), de-duplicated, and BPE-encoded (vocabulary 6000) into the
discrete tokens that are then fed to the LLM decoder.}
\label{fig:discrete-tokens-pipeline}
\end{figure}

This section describes the system used throughout the experiments. The system has two stages. First, a waveform is converted into a sequence of discrete tokens (Fig.~\ref{fig:discrete-tokens-pipeline}). Second, a LLM decoder transcribes these tokens into text. The tokenization stage follows the standard
discrete-token recipe~\cite{wang2025discrete,shon2024discreteslu}. Given a
raw waveform $\mathbf{x}$:

\begin{enumerate}
  \item \textbf{SSL feature extraction.} One or more SSL encoders
        extract frame-level continuous embeddings
        $\mathbf{H} = \{h_1, \ldots, h_T\}$.
  \item \textbf{K-means clustering.} Embeddings are mapped to
        discrete tokens $\mathbf{Z} = \{z_1, \ldots, z_T\}$ using
        K-means with $K = 2000$ centroids.
  \item \textbf{De-duplication.} Consecutive identical tokens are merged, yielding
$\tilde{\mathbf{Z}} = \{\tilde{z}_1, \ldots, \tilde{z}_{S}\}$
with $|\tilde{\mathbf{Z}}| \le |\mathbf{Z}|$, where $|\cdot|$ denotes
the number of tokens in a sequence.
  \item \textbf{BPE subword modeling.} Byte-pair encoding ($V = 6000$) compresses frequent subsequences into
        meta-tokens $\tilde{\mathbf{Z}}' = \{\tilde{z}'_1, \ldots, \tilde{z}'_{R}\}$.
\end{enumerate}

The token sequence $\tilde{\mathbf{Z}}'$ is then fed to the LLM decoder,
Qwen\,2.5-0.5B~\cite{bai2023qwen} with LoRA~\cite{hu2021lora}, which produces
the transcription autoregressively. Unless stated otherwise, all experiments
use $K = 2000$ centroids and a BPE vocabulary of 6000 units.

We use three SSL encoders to obtain complementary token views:
HuBERT-Large~\cite{hsu2021hubert}, WavLM-Large~\cite{chen2022wavlm}, and
MMS-300M~\cite{pratap2023mms}. HuBERT and WavLM are English SSL encoders with
similar architecture but different pre-training objectives, while MMS-300M is
multilingual and provides a more distant view. This gives us two close English
views and one more different multilingual view, which is useful for testing
whether encoder diversity improves discrete-token ASR.

For HuBERT-Large and WavLM-Large, we follow Wang et
al.~\cite{wang2025discrete} and extract the final layer, layer 24. For MMS-300M, we select layer 15 using a lightweight CTC probe: the
encoder is frozen, one CTC head is trained per layer, and the layer with the
lowest LibriSpeech test-clean WER is kept.

 

\section{Multi-Encoder Methods}
\label{sec:fusion}

We compare three strategies for leveraging multiple SSL encoders. The first two are
\emph{fusion baselines} that combine encoder outputs into a single
input representation before decoding. Early fusion operates at the embedding level before quantization
(Section~\ref{sec:early_fusion}), while late fusion operates at the token level after quantization
(Section~\ref{sec:late_fusion}). The third strategy is the proposed \emph{multi-view token augmentation}
method, which keeps encoder-specific tokenizations separate and trains a single
shared decoder across these views (Section~\ref{sec:shared_decoder}).

\subsection{Early Fusion (Embedding-Level, Pre-Quantization)}
\label{sec:early_fusion}

In early fusion, denoted by $\oplus$, the continuous embeddings of two SSL speech
encoders are concatenated frame-by-frame \emph{before} K-means
discretization:
\begin{equation}
  \mathbf{H}_{A \oplus B} =
    \bigl[\,\mathrm{Enc}_A(\mathbf{x}) \;\|\; \mathrm{Enc}_B(\mathbf{x})\,\bigr],
\label{eq:early_fusion}
\end{equation}
where $\|\,$ denotes feature-dimension concatenation. For two encoders with
1024-dimensional embeddings, this produces 2048-dimensional vectors. The
fused embeddings are then jointly discretized, de-duplicated, and BPE-encoded
into a single token stream. This asks K-means to learn one joint codebook over
the combined encoder space.

\subsection{Late Fusion (Token-Level, Post-Quantization)}
\label{sec:late_fusion}

In late fusion, each SSL encoder is discretized independently, producing
encoder-specific token sequences $\tilde{\mathbf{Z}}'^A$ and
$\tilde{\mathbf{Z}}'^B$. These sequences are generally of different lengths,
so fusion is performed after tokenization. We evaluate three variants.

\paragraph*{Tag-delimited fusion ($+_{\mathrm{tag}}$).}
Each stream is wrapped with encoder-specific tags added to the LLM
vocabulary:
\begin{equation}
  \tilde{\mathbf{Z}}'_{A +_{\mathrm{tag}} B}
  =
  [\texttt{<A>},\,\tilde{\mathbf{Z}}'^A,\,\texttt{</A>},
   \texttt{<B>},\,\tilde{\mathbf{Z}}'^B,\,\texttt{</B>}].
\label{eq:late_tag}
\end{equation}
This preserves encoder identity explicitly.

\paragraph*{Pipe-delimited fusion ($+_{\mathrm{pipe}}$).}
The two streams are concatenated with a single shared separator token, without
encoder-specific tags. This tests whether the decoder can infer encoder
identity from token statistics alone.

\paragraph*{String concatenation with joint BPE.} This strategy merges two
de-duplicated K-means streams before applying BPE. Since the two encoders may
produce sequences of different lengths, we first pair them by index up to the
shorter sequence. Each cluster ID is zero-padded to a fixed width of four digits,
and at every paired position the two padded IDs are concatenated into a single
symbol. Any remaining symbols from the longer sequence are appended unchanged.
The resulting sequence has the length of the longer stream and is used to train a
joint BPE tokenizer with a vocabulary size of 6000.


\subsection{Multi-View Token Augmentation (Proposed)}
\label{sec:shared_decoder}

Unlike the fusion baselines of Sections~\ref{sec:early_fusion}
and~\ref{sec:late_fusion}, the proposed multi-view token augmentation method
does not merge encoder outputs into a single input representation. Instead, it
treats each SSL encoder as an alternative discrete tokenizer of the same
utterance. The resulting encoder-specific token sequences are used as separate
training views for a single shared LLM decoder. The LLM
vocabulary is partitioned into three disjoint BPE blocks of 6000
tokens each, one per encoder. Each training sample is prefixed with
an encoder-identity tag
$\{\texttt{<hubert>}, \texttt{<wavlm>}, \texttt{<mms>}\}$, which
both selects the active vocabulary block and routes attention to the
correct token embeddings.

Because all parameters---LoRA weights and the language-model
head---are shared across encoders, each transcription is seen paired
with three different tokenizations of the same utterance. This is the
data-augmentation effect at the heart of our approach: the shared
parameters are regularized by alternative, equally valid views of every
training example, rather than by stochastic noise. The training
objective is the standard cross-entropy over text tokens only; the
encoder-identity tag is prepended to the audio prefix but excluded from
the loss.

At inference time the method supports two regimes. In the
\textbf{single-encoder} regime, the decoder is queried with the token
stream of one encoder alone, at exactly the cost of a single-encoder
baseline. In
the \textbf{multi-encoder} regime, the decoder is queried once per
encoder, producing three independent hypotheses at the cost of three
forward passes.

\section{Experimental Setup}
\label{sec:setup}
This section describes the experimental setup: the data and metrics
(Section~\ref{sec:data}), the decoder and training configuration
(Section~\ref{sec:train}) and ROVER post-processing
(Section~\ref{sec:rover}).




\subsection{Training and Evaluation Data}
\label{sec:data}

All models are trained on LibriSpeech 960h~\cite{panayotov2015librispeech}.
We use this corpus as a fixed training set for all systems in order to keep
the comparison controlled across single-encoder models, fusion baselines and
multi-view token augmentation. Our goal is not to
report state-of-the-art ASR results, but to isolate the effect of encoder
diversity under a common moderate-compute setup. We report in-domain results
on test-clean and test-other.

We also evaluate on Loquacious dev~\cite{parcollet2025loquacious}, which combines Common Voice~18.0, VoxPopuli, LibriSpeech dev/test-other, and YODAS. It contains broader English speech conditions than LibriSpeech and is used as a generalization test. Following the source-level diagnostic in Section~\ref{sec:res_yodas}, the main Loquacious dev results exclude YODAS. We mark these scores as Loq. dev$^\dagger$ in the tables.

We report Word Error Rate (WER) and Character Error Rate (CER). For the multi-view model we also report the \emph{oracle} WER, defined as the per-utterance minimum WER over the encoder hypotheses.

\subsection{Decoder, Training, and Decoding Configuration}
\label{sec:train}

We use Qwen\,2.5-0.5B~\cite{bai2023qwen} as the decoder and adapt it with
LoRA~\cite{hu2021lora} using $r{=}16$ and $\alpha{=}32$. All models are
trained for 5 epochs with AdamW, a learning rate of $2{\times}10^{-4}$, weight
decay $10^{-2}$, cosine scheduling, and an effective batch size of 16.
Training is performed on NVIDIA A100 or V100 GPUs depending on availability.

At inference time, we use beam-search decoding with 5 beams and
a repetition penalty of 1.2.

For multi-view token augmentation, the shared decoder is trained on $3N$
samples, with $N$ samples per encoder view, using separate BPE token blocks and
encoder tags.

 

\subsection{ROVER Post-Processing}
\label{sec:rover}

We use ROVER~\cite{fiscus1997rover}, through the NIST SCTK
implementation,\footnote{\url{https://github.com/usnistgov/SCTK}} as a
post-processing step to combine encoder-specific hypotheses. The encoders
and decoder are not updated. We report confidence-averaged voting
(\texttt{avgconf}), where each word vote is weighted by the average decoder
confidence of the systems proposing it. Word confidences are computed from
decoder token probabilities. When a word is split into several BPE tokens, we aggregate token probabilities with a geometric mean, which gives a
length-normalized word confidence and avoids penalizing words only because
they contain more subword tokens. Hyperparameters are tuned on
\texttt{dev-clean}$\cup$\texttt{dev-other} of LibriSpeech.

\section{Results}
\label{sec:results}

\subsection{Single-View Baselines}
\label{sec:results_single}

The first block of Table~\ref{tab:multiview} reports the single-encoder
baselines. The HuBERT system obtains the best result on LibriSpeech clean (3.71\% WER) and Loq.\ dev$^\dagger$ (21.96\% WER), while
WavLM is slightly better on LibriSpeech other (9.55\% vs.\ 9.70\%). MMS-300M is weaker
on both LibriSpeech splits, likely because its multilingual pre-training is
less specialized for English ASR. We note that on Loq.\ dev$^\dagger$, all encoders
degrade sharply, confirming that this set is a harder generalization test.

\begin{table}[!t]
    \centering
    \caption{Single-encoder baselines and multi-view token augmentation results.
    The shared-decoder multi-view model is trained on H, W, and M encoder-specific
token views and evaluated with one encoder at a time. Oracle selects
    the best hypothesis per utterance. $^\dagger$YODAS excluded. Best
    encoder row per block in bold. Values in \%.}
    \label{tab:multiview}
    \setlength{\tabcolsep}{3pt}
    \renewcommand{\arraystretch}{1.06}
    \scriptsize
    \resizebox{\columnwidth}{!}{%
    \begin{tabular}{lcccccc}
        \toprule
        \multirow{2}{*}{\textbf{Encoder}}
        & \multicolumn{2}{c}{\textbf{LS clean}}
        & \multicolumn{2}{c}{\textbf{LS other}}
        & \multicolumn{2}{c}{\textbf{Loq. dev$^\dagger$}} \\
        \cmidrule(lr){2-3}
        \cmidrule(lr){4-5}
        \cmidrule(lr){6-7}
        & \textbf{WER}$\downarrow$
        & \textbf{CER}$\downarrow$
        & \textbf{WER}$\downarrow$
        & \textbf{CER}$\downarrow$
        & \textbf{WER}$\downarrow$
        & \textbf{CER}$\downarrow$ \\
        \midrule

        \multicolumn{7}{l}{\emph{Single-encoder baselines}} \\
        \hspace{1em} HuBERT
        & \textbf{3.71} & \textbf{1.51}
        & 9.70 & 4.84
        & \textbf{21.96} & \textbf{12.70} \\

        \hspace{1em} WavLM
        & 4.08 & 1.81
        & \textbf{9.55} & \textbf{4.82}
        & 22.17 & 12.91 \\

        \hspace{1em} MMS-300M
        & 6.08 & 2.89
        & 15.10 & 8.34
        & 25.55 & 14.83 \\

        \hspace{1em} Oracle
        & 2.30 & 0.95
        & 6.68 & 3.48
        & 16.02 & 9.35 \\

        \midrule
        \multicolumn{7}{l}{\emph{Shared-decoder multi-view model}} \\
        \hspace{1em} HuBERT
        & 3.38 & 1.32
        & 8.37 & 4.00
        & 23.73 & 13.73 \\

        \hspace{1em} WavLM
        & \textbf{3.30} & \textbf{1.23}
        & \textbf{8.13} & \textbf{3.84}
        & \textbf{20.28} & \textbf{11.51} \\

        \hspace{1em} MMS-300M
        & 5.13 & 2.46
        & 13.29 & 7.38
        & 26.46 & 15.11 \\

        \hspace{1em} Oracle
        & 2.39 & 0.89
        & 5.96 & 2.88
        & 15.49 & 8.87 \\

        \bottomrule
    \end{tabular}%
    }
\end{table}


\subsection{Multi-View Token Augmentation}
\label{sec:res_multiview}

The second block of Table~\ref{tab:multiview} reports the performance of the
shared-decoder multi-view model when evaluated with each encoder view. On LibriSpeech, all
encoder views improve with multi-view training. WavLM gives the best decoded
view, reaching 3.30\% WER on LibriSpeech clean and 8.13\% on LibriSpeech other, compared with
4.08\% and 9.55\% for its independent baseline. This also outperforms the
best fusion baseline, HuBERT$\oplus$WavLM, while keeping single-encoder
inference cost.

The oracle rows show that the views remain complementary after training.
Oracle selection reaches 2.39\% WER on LS clean and 5.96\% on LS other, well
below any individual view. On Loq.\ dev$^\dagger$, the WavLM view improves
over its independent baseline (20.28\% vs.\ 22.17\% WER), and the oracle
drops further to 15.49\%. These results suggest that multi-view training both
improves the strongest view and preserves useful cross-encoder diversity.

\subsection{Comparison with Input-Level Fusion}
\label{sec:results_fusion}

\begin{table}[!h]
    \centering
  
    \caption{Fusion baselines using multiple SSL encoders. H, W, and M denote
    HuBERT-Large, WavLM-Large, and MMS-300M. $\oplus$ is early embedding
    fusion; $+_{\mathrm{tag}}$, $+_{\mathrm{pipe}}$, and $+_{\mathrm{str}}$
    are late token-fusion variants. All systems use Qwen\,2.5-0.5B with LoRA,
    $K$-means ($K{=}2000$), BPE 6000, and LS-960 training.
    $^\dagger$YODAS excluded. Best per column in bold. Values in \%.}
    \label{tab:fusion}

    \setlength{\tabcolsep}{2.3pt}
    \renewcommand{\arraystretch}{1.06}
    \scriptsize

    \resizebox{\columnwidth}{!}{%
    \begin{tabular}{lcccccc}
        \toprule
        \multirow{2}{*}{\textbf{Configuration}}
        & \multicolumn{2}{c}{\textbf{LS clean}}
        & \multicolumn{2}{c}{\textbf{LS other}}
        & \multicolumn{2}{c}{\textbf{Loq. dev$^\dagger$}} \\
        \cmidrule(lr){2-3}
        \cmidrule(lr){4-5}
        \cmidrule(lr){6-7}
        & \textbf{WER}$\downarrow$
        & \textbf{CER}$\downarrow$
        & \textbf{WER}$\downarrow$
        & \textbf{CER}$\downarrow$
        & \textbf{WER}$\downarrow$
        & \textbf{CER}$\downarrow$ \\
        \midrule

        \multicolumn{7}{l}{\emph{Best single-encoder baseline}} \\
        \hspace{1em} Best single enc.
        & 3.71 & 1.51
        & 9.55 & 4.82
        & \textbf{21.96} & \textbf{12.70} \\

        \midrule
        \multicolumn{7}{l}{\emph{Early fusion}} \\
        \hspace{1em} H $\oplus$ W
        & \textbf{3.55} & 1.57
        & \textbf{8.48} & \textbf{4.22}
        & 23.24 & 14.21 \\

        \hspace{1em} H $\oplus$ M
        & 4.97 & 2.19
        & 12.62 & 6.54
        & 24.29 & 13.68 \\

        \hspace{1em} W $\oplus$ M
        & 5.14 & 2.32
        & 13.61 & 7.06
        & 24.98 & 14.53 \\

        \hspace{1em} H $\oplus$ W $\oplus$ M
        & 6.94 & 3.41
        & 16.79 & 9.66
        & 29.74 & 17.43 \\

        \midrule
        \multicolumn{7}{l}{\emph{Late fusion: tag-delimited}} \\
        \hspace{1em} H $+_{\mathrm{tag}}$ W
        & 3.63 & \textbf{1.38}
        & 9.10 & 4.31
        & 24.08 & 14.65 \\

        \hspace{1em} H $+_{\mathrm{tag}}$ M
        & 4.90 & 2.18
        & 13.08 & 7.46
        & 36.13 & 21.49 \\

        \hspace{1em} W $+_{\mathrm{tag}}$ M
        & 3.70 & 1.51
        & 10.97 & 6.51
        & 25.73 & 15.96 \\

        \hspace{1em} H $+_{\mathrm{tag}}$ W $+_{\mathrm{tag}}$ M
        & 3.78 & 1.50
        & 9.62 & 4.85
        & 24.76 & 15.21 \\

        \midrule
        \multicolumn{7}{l}{\emph{Late fusion: pipe-delimited}} \\
        \hspace{1em} H $+_{\mathrm{pipe}}$ W
        & 3.77 & 1.55
        & 9.82 & 4.91
        & 28.48 & 18.77 \\

        \hspace{1em} H $+_{\mathrm{pipe}}$ M
        & 5.72 & 2.25
        & 9.65 & 4.73
        & 29.18 & 19.73 \\

        \hspace{1em} W $+_{\mathrm{pipe}}$ M
        & 4.03 & 1.68
        & 9.45 & 4.64
        & 22.04 & 13.09 \\

        \hspace{1em} H $+_{\mathrm{pipe}}$ W $+_{\mathrm{pipe}}$ M
        & 3.80 & 1.54
        & 12.05 & 7.35
        & 24.89 & 14.26 \\

        \midrule
        \multicolumn{7}{l}{\emph{Late fusion: string concatenation}} \\
        \hspace{1em} H $+_{\mathrm{str}}$ W
        & 4.16 & 1.83
        & 9.94 & 4.98
        & 22.62 & 13.06 \\

        \hspace{1em} H $+_{\mathrm{str}}$ M
        & 4.47 & 2.07
        & 11.95 & 7.13
        & 28.14 & 18.98 \\

        \hspace{1em} W $+_{\mathrm{str}}$ M
        & 4.93 & 2.75
        & 11.21 & 6.57
        & 29.92 & 20.22 \\

        \hspace{1em} H $+_{\mathrm{str}}$ W $+_{\mathrm{str}}$ M
        & 4.48 & 2.07
        & 11.40 & 6.44
        & 30.64 & 20.27 \\

        \bottomrule
    \end{tabular}%
    }
\end{table}

Table~\ref{tab:fusion} compares fusion baselines that merge several encoder
streams before decoding. Changing the encoder order led to only minor
variations and did not change the overall conclusion, so we report one order
for compactness. Early fusion is the strongest in-domain method:
HuBERT$\oplus$WavLM reaches 3.55\% WER on test-clean and 8.48\% on
test-other, improving over the best single-encoder baselines. Late fusion is
less stable. The tag-delimited HuBERT+WavLM system is competitive on
test-clean, but it is weaker than early fusion on test-other, while pipe and
string concatenation give mixed results. Adding MMS-300M generally degrades input-level fusion performance, suggesting that its different token view is not easy to
exploit once streams are collapsed into one input.

On Loq.\ dev$^\dagger$, fusion gives limited generalization gains. The best
fusion result is WavLM+MMS with pipe delimitation at 22.04\% WER and
13.09\% CER, which is close to but still worse in WER than the best
single-encoder baseline. Thus, input-level fusion can help on LibriSpeech,
but it does not provide a robust solution under broader speech conditions.
This motivates our multi-view token augmentation strategy, which keeps
encoder-specific token views separate instead of merging them before decoding.

\subsection{Loquacious Source Analysis}
\label{sec:res_yodas}

The full Loquacious dev split initially produced unusually high aggregate
WER. To identify the source, we evaluated each source subset separately.
Table~\ref{tab:loquacious_origin_dev} shows that Common Voice, LibriSpeech,
and VoxPopuli follow expected degradation patterns, while YODAS diverges
strongly. In the shared-decoder multi-view model, YODAS reaches 95--121\%
WER, indicating a severe mismatch rather than a normal robustness drop. We
therefore exclude YODAS from the main Loq.\ dev$^\dagger$ scores, while
reporting it here to make the exclusion explicit.

\begin{table}[t]
    \centering
    \caption{Loquacious dev WER/CER (\%) by source corpus. H, W, and M denote
    HuBERT-Large, WavLM-Large, and MMS-300M. Each cell reports WER/CER.
    YODAS diverges strongly and is excluded from the main Loq.\ dev$^\dagger$
    scores.}
    \label{tab:loquacious_origin_dev}

    \setlength{\tabcolsep}{3pt}
    \renewcommand{\arraystretch}{1.08}
    \scriptsize

    \resizebox{\columnwidth}{!}{%
    \begin{tabular}{lcccc}
        \toprule
        \textbf{System}
        & \textbf{Common Voice} 
        & \textbf{LibriSpeech}
        & \textbf{VoxPopuli}
        & \textbf{YODAS} \\
        \midrule

        \multicolumn{5}{l}{\emph{Single-encoder baselines}} \\
        \hspace{0.6em} H
        & 37.34 / 21.74
        & 9.27 / \textbf{4.73}
        & 26.13 / \textbf{14.90}
        & \textbf{51.39} / \textbf{33.72} \\

        \hspace{0.6em} W
        & 39.00 / 22.41
        & \textbf{9.12} / 4.79
        & \textbf{25.78} / 14.97
        & 64.26 / 45.17 \\

        \hspace{0.6em} M
        & \textbf{37.12} / \textbf{20.62}
        & 13.92 / 7.48
        & 31.07 / 18.55
        & 65.51 / 42.67 \\

        \midrule
        \multicolumn{5}{l}{\emph{Shared-decoder multi-view model}} \\
        \hspace{0.6em} H
        & 47.59 / 27.78
        & \textbf{8.13} / \textbf{4.03}
        & 24.96 / 14.70
        & 121.17 / 77.53 \\

        \hspace{0.6em} W
        & \textbf{37.59} / \textbf{21.49}
        & 8.23 / 4.52
        & \textbf{22.32} / \textbf{12.08}
        & \textbf{95.47} / \textbf{61.20} \\

        \hspace{0.6em} M
        & 41.44 / 22.50
        & 11.83 / 6.27
        & 32.71 / 18.98
        & 117.82 / 77.97 \\

        \bottomrule
    \end{tabular}%
    }
\end{table}

\subsection{Isolating the Role of Encoder Diversity}
\label{sec:res_diversity}

We test whether the gain comes from encoder diversity or simply from showing
the decoder more token sequences. Table~\ref{tab:ablation_data} compares
fixed-budget multi-view training with a 3-stream control where the same
encoder is repeated three times. Repetition does not reproduce the full
multi-view gain: WavLM-3x reaches 4.99\% WER on LibriSpeech clean, while the full
multi-view WavLM view reaches 3.30\%. The fixed-budget settings are also
weaker than the full model, showing that each view still needs enough data.
However, their oracle rows remain strong, especially on Loq.\ dev$^\dagger$,
where they reach about 17.8\% WER. This shows that the views still make
complementary errors under a reduced budget.

Table~\ref{tab:oracle_pairwise} confirms the same trend with two-encoder
shared decoders. The best pair is not always the pair with the strongest
single-encoder baselines, and pairs involving MMS-300M can improve the
stronger English encoder view despite MMS-300M being weaker alone. Overall,
the gain comes from complementary tokenizations, not from adding more copies
of the same representation. This remaining complementarity motivates the
ROVER experiments in Section~\ref{sec:results_rover}.

\begin{table}[t]
    \centering
    \caption{Data-budget and repetition controls. H/W/M denote
    HuBERT-Large, WavLM-Large, and MMS-300M. Same-$N/3$ uses the same
    utterances across encoders; split-$N/3$ uses different utterances per
    encoder. The 3-stream control repeats one encoder three times.
    $^\dagger$YODAS excluded. Best per block and column in bold. Values in \%.}
    \label{tab:ablation_data}
    \setlength{\tabcolsep}{2.5pt}
    \renewcommand{\arraystretch}{1.06}
    \scriptsize
    \resizebox{\columnwidth}{!}{%
    \begin{tabular}{lcccccc}
        \toprule
        \multirow{2}{*}{\textbf{Model}}
        & \multicolumn{2}{c}{\textbf{LS clean}}
        & \multicolumn{2}{c}{\textbf{LS other}}
        & \multicolumn{2}{c}{\textbf{Loq. dev$^\dagger$}} \\
        \cmidrule(lr){2-3}
        \cmidrule(lr){4-5}
        \cmidrule(lr){6-7}
        & \textbf{WER}$\downarrow$
        & \textbf{CER}$\downarrow$
        & \textbf{WER}$\downarrow$
        & \textbf{CER}$\downarrow$
        & \textbf{WER}$\downarrow$
        & \textbf{CER}$\downarrow$ \\
        \midrule

        \multicolumn{7}{l}{\emph{Same-$N/3$ utterances across H/W/M}} \\
        \hspace{1em} H
        & 4.57 & 1.88
        & 12.11 & 6.52
        & 31.22 & 19.89 \\
        \hspace{1em} W
        & 7.28 & 4.03
        & 10.35 & 5.92
        & 29.80 & 18.91 \\
        \hspace{1em} M
        & 6.82 & 3.31
        & 16.17 & 8.59
        & 31.80 & 19.90 \\
        \hspace{1em} Oracle
        & \textbf{3.16} & \textbf{1.30}
        & \textbf{7.77} & \textbf{3.90}
        & \textbf{17.75} & \textbf{10.32} \\

        \midrule
        \multicolumn{7}{l}{\emph{Split-$N/3$ utterances across H/W/M}} \\
        \hspace{1em} H
        & 4.64 & 1.85
        & 10.66 & 5.30
        & 27.96 & 16.74 \\
        \hspace{1em} W
        & 4.66 & 1.93
        & 11.36 & 6.03
        & 24.72 & 14.32 \\
        \hspace{1em} M
        & 7.50 & 3.76
        & 18.12 & 9.50
        & 29.00 & 17.32 \\
        \hspace{1em} Oracle
        & \textbf{3.19} & \textbf{1.29}
        & \textbf{7.83} & \textbf{3.89}
        & \textbf{17.81} & \textbf{10.30} \\

        \midrule
        \multicolumn{7}{l}{\emph{Single-encoder 3-stream control}} \\
        \hspace{1em} H-3x
        & \textbf{3.88} & \textbf{1.61}
        & 9.66 & \textbf{4.78}
        & 24.84 & 15.46 \\
        \hspace{1em} W-3x
        & 4.99 & 2.69
        & \textbf{9.53} & 4.90
        & \textbf{22.42} & \textbf{12.97} \\
        \hspace{1em} M-3x
        & 8.06 & 4.55
        & 14.03 & 7.82
        & 30.10 & 18.78 \\

        \bottomrule
    \end{tabular}%
    }
\end{table}

\begin{table}[t]
    \centering
    \caption{Pairwise shared-decoder results. H/W/M denote HuBERT-Large,
    WavLM-Large, and MMS-300M. Oracle selects the best hypothesis per
    utterance from the two encoder views. $^\dagger$YODAS excluded.
    Best oracle result per column in bold. Values in \%.}
    \label{tab:oracle_pairwise}
    \setlength{\tabcolsep}{2.5pt}
    \renewcommand{\arraystretch}{1.06}
    \scriptsize
    \resizebox{\columnwidth}{!}{%
    \begin{tabular}{llcccccc}
        \toprule
        \multirow{2}{*}{\textbf{Pair}}
        & \multirow{2}{*}{\textbf{View}}
        & \multicolumn{2}{c}{\textbf{LS clean}}
        & \multicolumn{2}{c}{\textbf{LS other}}
        & \multicolumn{2}{c}{\textbf{Loq. dev$^\dagger$}} \\
        \cmidrule(lr){3-4}
        \cmidrule(lr){5-6}
        \cmidrule(lr){7-8}
        & & \textbf{WER}$\downarrow$ & \textbf{CER}$\downarrow$
          & \textbf{WER}$\downarrow$ & \textbf{CER}$\downarrow$
          & \textbf{WER}$\downarrow$ & \textbf{CER}$\downarrow$ \\
        \midrule

        \multirow{3}{*}{H+M}
        & H
        & 3.45 & 1.30
        & 8.74 & 4.23
        & 22.79 & 13.42 \\
        & M
        & 5.13 & 2.37
        & 12.92 & 6.75
        & 26.06 & 15.14 \\
        & Oracle
        & \textbf{2.82} & \textbf{1.06}
        & 7.53 & 3.70
        & 17.24 & 9.63 \\

        \midrule
        \multirow{3}{*}{H+W}
        & H
        & 3.72 & 1.66
        & 9.28 & 4.71
        & 29.90 & 18.31 \\
        & W
        & 6.04 & 3.74
        & 8.97 & 4.55
        & 25.28 & 15.83 \\
        & Oracle
        & 2.90 & 1.28
        & \textbf{7.14} & 3.66
        & 18.26 & 10.74 \\

        \midrule
        \multirow{3}{*}{W+M}
        & W
        & 3.52 & 1.39
        & 8.26 & 3.91
        & 20.83 & 11.97 \\
        & M
        & 5.23 & 2.51
        & 12.56 & 6.63
        & 24.57 & 14.69 \\
        & Oracle
        & 2.84 & 1.13
        & 7.19 & \textbf{3.47}
        & \textbf{16.40} & \textbf{9.13} \\

        \bottomrule
    \end{tabular}%
    }
\end{table}

\subsection{Combining Multi-View Hypotheses with ROVER}
\label{sec:results_rover}

Table~\ref{tab:rover} tests whether the remaining complementarity between
encoder views can be used without oracle selection. We report
confidence-averaged ROVER (\texttt{avgconf}) over the three hypotheses
H+W+M. For the single-encoder hypotheses, ROVER improves WER from 3.71 to
3.14 on LibriSpeech clean and from 9.55 to 7.88 on LibriSpeech other. For the shared-decoder
multi-view hypotheses, ROVER gives the best LibriSpeech results of the paper,
reaching 3.03\% WER on LibriSpeech clean and 7.38\% on LibriSpeech other. On Loq.\
dev$^\dagger$, ROVER also improves over the best shared-decoder view in CER,
but the single-encoder ROVER system gives the lowest WER. These results show
that output-level combination recovers part of the residual complementarity
left by multi-view training.

\begin{table}[t]
    \centering
    \caption{Output-level combination with confidence-averaged ROVER
    (\texttt{avgconf}). H/W/M denote HuBERT-Large, WavLM-Large, and
    MMS-300M. $^\dagger$YODAS excluded. Best ROVER result per column in
    bold. Values in \%.}
    \label{tab:rover}
    \setlength{\tabcolsep}{3pt}
    \renewcommand{\arraystretch}{1.06}
    \scriptsize
    \resizebox{\columnwidth}{!}{%
    \begin{tabular}{lcccccc}
        \toprule
        \multirow{2}{*}{\textbf{System}}
        & \multicolumn{2}{c}{\textbf{LS clean}}
        & \multicolumn{2}{c}{\textbf{LS other}}
        & \multicolumn{2}{c}{\textbf{Loq. dev$^\dagger$}} \\
        \cmidrule(lr){2-3}
        \cmidrule(lr){4-5}
        \cmidrule(lr){6-7}
        & \textbf{WER}$\downarrow$
        & \textbf{CER}$\downarrow$
        & \textbf{WER}$\downarrow$
        & \textbf{CER}$\downarrow$
        & \textbf{WER}$\downarrow$
        & \textbf{CER}$\downarrow$ \\
        \midrule

        \multicolumn{7}{l}{\emph{Single-encoder hypotheses}} \\
        Best single view
        & 3.71 & 1.51
        & 9.55 & 4.82
        & 21.96 & 12.70 \\
        ROVER H+W+M
        & 3.14 & 1.52
        & 7.88 & 4.54
        & \textbf{17.39} & 11.18 \\

        \midrule
        \multicolumn{7}{l}{\emph{Shared-decoder multi-view hypotheses}} \\
        Best decoded view
        & 3.30 & 1.23
        & 8.13 & 3.84
        & 20.28 & 11.51 \\
        ROVER H+W+M
        & \textbf{3.03} & \textbf{1.26}
        & \textbf{7.38} & \textbf{3.73}
        & 18.22 & \textbf{10.63} \\

        \bottomrule
    \end{tabular}%
    }
\end{table}

\section{Discussion}
\label{sec:discussion}

This section discusses why the method works (Section~\ref{sec:disc_why}), how it can be used (Section~\ref{sec:disc_usage}), and its limitations and next steps (Section~\ref{sec:disc_limits}).

\subsection{Why Does It Work?}
\label{sec:disc_why}

The results suggest that multi-view token augmentation acts as a form of
training-time augmentation in the discrete SSL-token space. Each encoder gives a different tokenization of the
same utterance, and the shared decoder learns to map all views to the same
transcript. This prevents the decoder from depending too strongly on the
biases of one tokenizer and makes the mapping from speech tokens to text more
robust.

The ablations show that diversity is the key factor. Repeating the same
encoder three times does not reproduce the gain, even though the decoder sees
the same number of training streams. Pairwise results also show that the best
pairs are not always the strongest individual encoders. A weaker encoder such
as MMS-300M can still help when it makes different errors. This explains why
multi-view training is more effective than simply adding more copies of the
same representation.

These results also clarify the limits of input-level fusion. Early fusion can
help on LibriSpeech, but it collapses the encoder views before decoding. Late
fusion is also unstable. In contrast, multi-view token augmentation keeps the encoder-specific views
separate during training, and ROVER keeps them separate at the hypothesis
level. This makes it easier to exploit complementary errors between encoders.

\subsection{How Can It Be Used?}
\label{sec:disc_usage}

The method supports several inference regimes. The simplest one is
single-encoder inference: the decoder is trained with several token views, but
only one encoder is used at test time. This has the same inference cost as a
standard single-encoder system and is useful when latency or compute is
limited.

A second regime is multi-hypothesis decoding. When more compute is available,
the three encoders can be run in parallel and their outputs combined with
ROVER. This gives the best LibriSpeech results in our experiments, showing
that useful complementarity remains after multi-view training.

A third possibility is encoder selection. Since the same decoder can consume
tokens from any trained encoder, a future system could choose the most
reliable encoder for each utterance. The oracle results show that this could
bring large gains, because many errors are view-specific. A learned router
could select an encoder before decoding, or select among hypotheses after
decoding.

\subsection{Limitations and Next Steps}
\label{sec:disc_limits}

This study is a controlled proof of concept rather than a state-of-the-art ASR
system. Training only on LibriSpeech 960h keeps the comparison clean, but
limits the conclusions to read audiobook speech. Future work should therefore
train on more diverse English data, including conversational, noisy,
spontaneous, and web-sourced speech.

We also focus on discrete SSL-token pipelines, since our goal is to improve
discrete-token LLM-based ASR. A natural next step is to test whether the same
multi-view principle extends to continuous SSL embeddings, where different
encoders provide alternative continuous views.

Finally, Loquacious YODAS remains a clear failure case, suggesting a strong
domain mismatch with LibriSpeech-only training. This motivates both more
diverse training data and adaptive combination methods, such as learned encoder
routing, confidence-aware fusion, or mixture-of-experts decoders.

\section{Conclusion}
\label{sec:conclusion}

We introduced \emph{multi-view discrete token augmentation} for LLM-based ASR.
Rather than fusing SSL encoders into a single representation, we treat fixed
encoders as alternative tokenizers of the same utterance and train one shared
decoder across these views. At inference time, the model can use either a
single encoder, keeping standard inference cost, or multiple encoder hypotheses
combined at the output level.

On LibriSpeech, multi-view training improves all encoder views over their
single-encoder baselines. The best single-view inference result is obtained
with WavLM, reaching 3.30\% WER on test-clean and 8.13\% on test-other, while
confidence-averaged ROVER further improves performance to 3.03\% and 7.38\%.
Ablations show that these gains come from complementary encoder views rather
than data volume alone, since repeating one encoder does not reproduce the
effect.

These results suggest that SSL encoders are more effective when used as
distinct token views than when collapsed into a single fused input. Future work
should test the same principle with continuous SSL embeddings, train on more
diverse speech sources, and explore routing or mixture-of-experts mechanisms to
select the most reliable view for each utterance or domain.

\clearpage

\section*{Acknowledgment}

This work was granted access to the HPC resources of
IDRIS under the allocation AD010617089 made by
GENCI. EDL financially supported this work.


\end{document}